\documentclass[aps,prl,reprint,superscriptaddress,nofootinbib]{revtex4-2}

\usepackage{amsmath,amssymb,bm}
\usepackage{xfrac}
\usepackage{bbold}
\usepackage{graphicx}
\usepackage{hyperref}

\begin{document}

\title{Few-Electron Partitioning in the Fractional Quantum Hall Regime}

\begin{abstract}
Motivated by few-electron partitioning experiments [Shaju \textit{et al.}, Nature \textbf{642}, 928 (2025)], we study the partition statistics of $N$-electron states at a fractional quantum Hall point contact. At the solvable point $g=1/2$, we obtain nonperturbative current and noise formulas for arbitrary drives, while for Lorentzian pulses (Levitons) the full counting statistics reduces to an $N\times N$ determinant. We show that the charge is transferred solely in units of $e$, and that symmetrized cumulants for $N=2,3,4$ reveal strong antibunching and higher-order correlations, observable in a $(\nu_1,\nu_2)=(1,1/3)$ junction.
\end{abstract}

\author{T. Jonckheere}
\affiliation{Aix Marseille Univ, Universit\'e de Toulon, CNRS, CPT, Marseille, France}
\email{thibaut.jonckheere@cpt.univ-mrs.fr }
\author{J. Rech}
\affiliation{Aix Marseille Univ, Universit\'e de Toulon, CNRS, CPT, Marseille, France}
\author{B. Gr\'emaud}
\affiliation{Aix Marseille Univ, Universit\'e de Toulon, CNRS, CPT, Marseille, France}

\maketitle

\textit{Introduction.}--- Partitioning experiments at a quantum point contact (QPC) provide a remarkably direct window onto the microscopic organization and correlations of quantum states~\cite{Blanter2000,Bocquillon2014,Bauerle2018}. In the integer quantum Hall and ballistic regimes, on-demand electron sources---ranging from mesoscopic capacitors and dynamical quantum dots to Lorentzian voltage pulses (Levitons)~\cite{Levitov1993,Keeling2006,Feve2007,Dubois2013,Jullien2014,Fletcher2023}---have enabled the controlled partitioning of few-electron states. 
In 2015, Ubbelohde \textit{et al.}~\cite{Ubbelohde2015} reconstructed the partition probabilities of two-electron states from current and noise measurements. Very recently, Shaju \textit{et al.}~\cite{Shaju2025} used the full partitioning statistics of droplets containing only $N \le 5$ electrons to resolve interaction-induced correlations, including genuine all-body correlations, and to identify signatures of a strongly correlated Coulomb liquid. This establishes few-electron partitioning as a sensitive probe of correlations well before the thermodynamic limit is reached.

One may wonder how this picture changes when the host medium itself is a strongly correlated quantum fluid. The fractional quantum Hall (FQH) regime is particularly suited to address this question, as its low-energy edge excitations are collective and intrinsically correlated. In the FQH regime, stationary QPC partitioning of dilute beams has famously revealed fractional quasiparticle charges through shot noise~\cite{dePicciotto1997,Saminadayar1997} and, more recently, anyonic braiding statistics in collider geometries~\cite{Bartolomei2020,Nakamura2020,Glidic2023,Ruelle2023,Lee2023}, while time-dependent Lorentzian drives have been proposed and explored to generate minimal fractional excitations~\cite{Rech2017,Bisognin2019}. Yet, much less is known about how the intrinsic correlations of an FQH edge are encoded in the full partition statistics of a controlled few-electron excitation, and whether departures from independent-electron partitioning can be resolved at the level of a small, fixed number $N$ of injected electrons.
\begin{figure}[t]
\centering
\includegraphics[width=7.5cm]{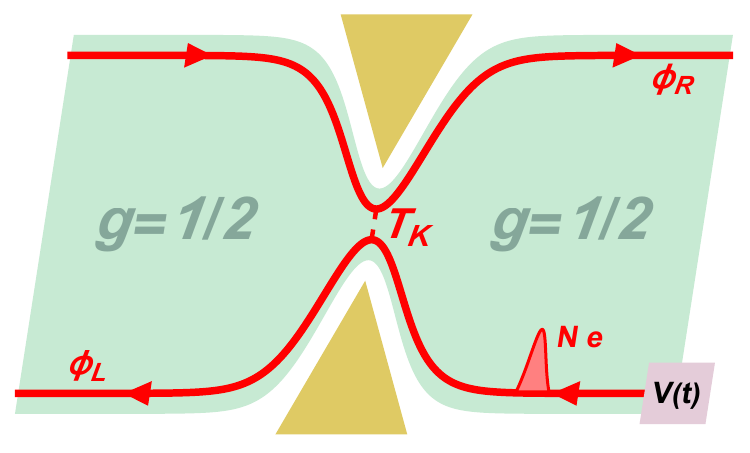}
\caption{Effective homogeneous setup at Luttinger parameter $g=1/2$: an integer Lorentzian voltage pulse $V(t)$ applied on the incoming $\phi_L$ edge injects $N$ electrons (Levitons) toward a QPC, where they are partitioned into the outgoing edges.}
\label{fig:setup}
\end{figure}

Here we answer this question nonperturbatively for a driven quantum point contact at the exactly solvable Luttinger parameter $g=1/2$. 
We focus throughout on states containing an integer number $N$ of injected electrons, generated with integer-charge Lorentzian pulses (Levitons).
We first derive exact formulas for the first two cumulants (the mean transferred charge and its fluctuations), valid for an arbitrary voltage pulse, and show that charge transfer at the QPC always occurs in units of $e$. We then develop an independent full-counting-statistics (FCS) approach, which reduces to the evaluation of a finite $N \times N$ matrix. The resulting partition probabilities are not those of $N$ independent trials: they contain genuine few-electron correlations specific to the strongly correlated system, as directly revealed by symmetrized multivariate cumulants.

The minimal nontrivial case, $N=2$, is already experimentally relevant: we show that the average transferred charge and its fluctuations are sufficient to reveal the electronic correlations~\cite{Ubbelohde2015}. For $N>2$, measurements of higher-order cumulants, or ultimately of the full counting statistics, would provide access to increasingly detailed few-electron correlations. Importantly, the effective $g=1/2$ results are directly observable in a physical inhomogeneous 
$(\nu_1,\nu_2)=(1,\sfrac{1}{3})$ junction, which maps exactly onto the $g=1/2$ impurity problem~\cite{Sandler1999}. We discuss at the end of this letter how this provides a concrete experimental route to the partitioning physics discussed here.

\textit{Model and exact solution.---}
We consider edge states of the FQHE at the effective Luttinger parameter $g=1/2$, with a quantum point contact connecting the two counter-propagating edge states [Fig.~\ref{fig:setup}]. We choose a gauge in which the scatterer is stationary and the voltage drive is encoded in the incoming state. 
The Hamiltonian of the left and right moving edge states can be expressed with the bosonic fields $\phi_L$ and $\phi_R$:
$H_0=\frac{v_F}{4\pi} \int dx\,[\partial_x\phi_L(x)]^2 + [\partial_x\phi_R(x)]^2$ (with $v_F$ the propagation velocity along the edge),
 and the tunneling at the QPC is written as
$H_T = \frac{1}{2\pi a} \Gamma e^{i \sqrt{g} (\phi_L(0) - \phi_R(0))}+ \mbox{H.c.}$, where $\Gamma$ is the tunneling amplitude, and $a$ is the short-distance
cutoff. Using $g=\sfrac{1}{2}$, and introducing
the symmetric and antisymmetric combination of the incoming edge states $\phi_{\pm}=\frac{1}{\sqrt{2}} (\phi_L \pm \phi_R)$  gives $H = H_0 + H_T$, with:
\begin{align}
H_0&=\frac{v_F}{4\pi}\sum_{r=\pm}\int dx\,[\partial_x\phi_r(x)]^2,\label{eq:H0}\\
H_T&=\frac{1}{2\pi a}\left[ \Gamma e^{i\phi_-(0)}+ \Gamma^* e^{-i\phi_-(0)}\right],\label{eq:HT}
\end{align}
so that the \(+\) mode decouples from the point contact. The vertex operator $e^{i\phi_-(0)}$ in Eq.~(\ref{eq:HT}) is identical to a fermion, reducing the interacting point contact to a quadratic boundary problem~\cite{Sandler1999}. The derivation is summarized in the End Matter. In terms of the crossover scale
 $T_K  \propto |\Gamma|^2$, the exact relation describing the scattering of the incoming fermion $\psi_{\rm in}$ 
into the outgoing fermion $\psi_{\rm out}$ at the QPC is
\begin{equation}
\psi_{\rm out}(\omega)
=
\frac{\omega}{\omega+iT_K}\psi_{\rm in}(\omega)
+
\frac{iT_K}{\omega+iT_K}\psi_{\rm in}^{\dagger}(-\omega).
\label{eq:scattering}
\end{equation}
The two terms correspond to normal transmission of an incoming
fermion and Andreev reflection into a hole, respectively.
The voltage pulse $V(t)$ is entirely encoded in the incoming refermionized field,
\begin{equation}
\psi_{\rm in}(t)=e^{-i\varphi(t)}\psi^{(0)}_{\rm in}(t),
\quad
\varphi(t)=\frac{e^*}{\hbar}\int_{-\infty}^{t}dt'\,V(t'),
\label{eq:drive}
\end{equation}
with $e^*=\frac{e}{2}$.
We define the driven incoming Green function as
\begin{align}
G_V(t,t')&\equiv
\left\langle\psi_{\rm in}^{\dagger}(t')\psi_{\rm in}(t)\right\rangle_V
\nonumber\\
&=e^{-i[\varphi(t)-\varphi(t')]}G_0(t-t'),
\label{eq:drivencorr}
\end{align}
where \(G_0(t-t')\equiv\langle\psi_{\rm in}^{(0)\dagger}(t')\psi_{\rm in}^{(0)}(t)\rangle\); at zero temperature and with the short-time cutoff $t_0$, 
$G_0(\tau)=[2\pi(t_0+i\tau)]^{-1}$. 
%Thus the scatterer remains stationary for an arbitrary \(V(t)\), and all drive dependence enters through the phase differences in Eq.~(\ref{eq:drivencorr}).

We compute the transferred charge in units of \(e\),
\begin{equation}
Q=\frac{1}{2}\int dt\,
\left[
\psi_{\rm in}^{\dagger}(t)\psi_{\rm in}(t)
-\psi_{\rm out}^{\dagger}(t)\psi_{\rm out}(t)
\right].
\label{eq:charge}
\end{equation}
Using Eqs.~(\ref{eq:scattering}) and (\ref{eq:drivencorr}) gives, for an arbitrary pulse,
\begin{align}
\left \langle I(t) \right \rangle=
\frac{e T_K}{2\pi} &
\int_{-\infty}^{t}dt'\,
e^{-T_K(t-t')}
\sin\!\big[\varphi(t)-\varphi(t')\big] \nonumber \\
&\times \frac{t-t'}{(t-t')^2+t_0^2},
\label{eq:Iarbitrary}
\end{align}
and
\begin{equation}
C_1=\langle Q\rangle=\frac{1}{e} \int dt\,\left \langle I(t) \right \rangle,
\label{eq:C1}
\end{equation} where we introduce the notation $C_{i}$ for the $i^{th}$ cumulant
of the charge. The charge fluctuations follow from
\begin{equation}
C_2=\langle Q^2\rangle-\langle Q\rangle^2
=\frac{1}{e^2} \int dt_1dt_2\,\langle\delta I(t_1)\delta I(t_2)\rangle,
\label{eq:C2}
\end{equation}
where $\delta I(t) = I(t) - \langle I(t) \rangle$.
The explicit average appearing in Eq.(\ref{eq:C2}) can be obtained by applying Wick theorem, and is given in the Eq.(\ref{eq:Sexplicit}) of the End Matter.
%Since the incoming state is Gaussian, the four-point functions entering Eq.~(\ref{eq:C2}) reduce by Wick's theorem.
Equations~(\ref{eq:Iarbitrary}) and (\ref{eq:C2}) thus provide a nonperturbative calculation of the first two cumulants for an arbitrary voltage pulse.
% the explicit noise kernel and the intermediate refermionization steps are given in the End Matter.

\textit{Few-electron partitioning.---}
\begin{figure}[t]
\centering
\includegraphics[width=\columnwidth]{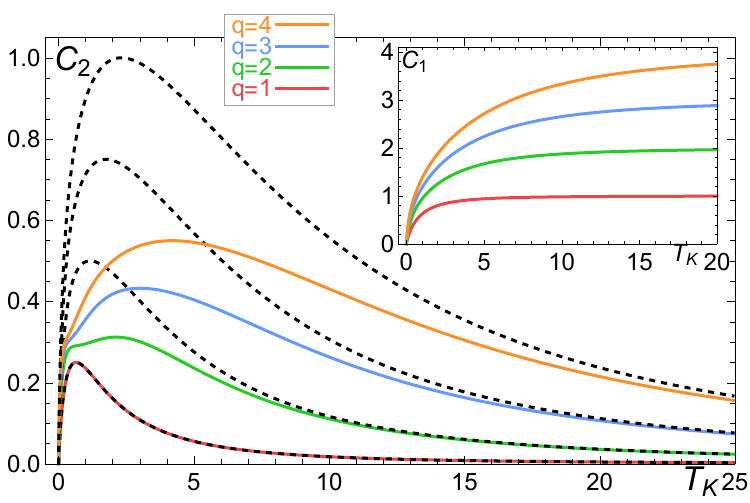}
\caption{Charge fluctuations
$C_2=\langle Q^2\rangle-\langle Q\rangle^2$ as a function
of the QPC scale $T_K$, for Leviton states containing
$q=1,2,3,4$ electrons (from bottom to top), obtained from the direct integration of
Eq.~(\ref{eq:C2}), or equivalently from the FCS
Eq.~(\ref{eq:FCSC2}). Black dashed lines show the binomial result  $C_2^{\rm bin}$ of Eq.(\ref{eq:binomial}) for
$q$ independently partitioned electrons.
Inset: average transferred charge $C_1=\langle Q\rangle$ as a function of $T_K$ for $q=1,2,3,4$ (from bottom to top), approaching $C_1=q$ in the large-$T_K$ limit.}
\label{fig:C2all}
\end{figure}
%Eqs.~(\ref{eq:Iarbitrary})-(\ref{eq:C2}) hold for an arbitrary drive.
 As our goal is to study the partitioning of a state with a few electrons, we now
focus on integer-charge Lorentzian pulses,
\begin{equation}
V(t)=\frac{2q\hbar}{e^*}\,
\frac{\tau_0}{t^2+\tau_0^2},
\qquad
\frac{e^*}{h}\int_{-\infty}^{\infty}dt\,V(t)=N
\label{eq:Lorentzianshape}
\end{equation} 
where $\tau_0$ is the temporal width of the pulse. These pulses have the property, when $q=N$ is an integer, to create a state
with $N$ electrons
without additional electron-hole excitations, and have been studied theoretically and experimentally~\cite{Levitov1993,Keeling2006,Dubois2013,Grenier2013,Rech2017}.
In the following, we use $\tau_0=1$, and $T_K$ will be considered as the dimensionless quantity $T_K \tau_0$.
 Fig.~\ref{fig:C2all} shows the charge fluctuations $C_2$ for $q=1,2,3,4$, obtained from the direct integration of
Eq.~(\ref{eq:C2}) with the pulse shape given in Eq.~\eqref{eq:Lorentzianshape}. As a reference, we consider $q$ independent electrons,
each transmitted with probability $p=C_1/q$, with $C_1$ computed from Eq.(\ref{eq:C1}). Their
binomial fluctuations are
\begin{equation}
C_2^{\rm bin}=q p(1-p)
=C_1\left(1-\frac{C_1}{q}\right).
\label{eq:binomial}
\end{equation}
which are shown as dashed lines on Fig.~\ref{fig:C2all}.
For $q=1$, the exact result coincides with the binomial
one for all $T_K$. For $q\geq2$, the fluctuations are
systematically reduced with respect to
Eq.~(\ref{eq:binomial}), with a $T_K$ dependence quite different for intermediate $T_K$, and the two results coinciding
 in the $T_K\to0$ and $T_K\to\infty$ limits.
 As it turns out, $q=2$ is already sufficient to observe a marked departure from binomial fluctuations.

The $q=1$ result suggests a purely electronic partition,
despite the charge-$e/2$ excitations of the effective
$g=1/2$ theory. 
This is a consequence of the $\phi_\pm$ decomposition and the scattering relation of Eq.~(\ref{eq:scattering}):
an injected electron splits into an incoming charge $e/2$ in the decoupled $\phi_+$ mode and 
an incoming charge-$e/2$ excitation in the $\phi_-$ mode, which is either normally scattered ($+e/2$)
 or Andreev reflected into a hole ($-e/2$), recombining into a transferred charge $0$ or $e$, respectively~\cite{KaneFisher2003}.
This does not contradict the effective charge $e/2$
inferred from DC shot noise, as obtained through a calculation in the refermionized formalism in the DC regime~\cite{Chamon1995,Sandler1999,KaneFisher2003}.
This point has already been discussed by Andreev et al. in   Ref.~\onlinecite{AndreevMishchenko} for a related system:
in the dilute, long-time limit
the global parity constraint requiring an even number of
charge-$e/2$ transfer events becomes irrelevant for the
lowest cumulants, which then coincide with those of a
Poisson process of charge-$e/2$ quasiparticles.

To make the purely electronic partition statement quantitative, we allow for the
three transfer sectors $0,e/2,e$, with probabilities $P_{20}$,
$P_{11}$, and $P_{02}$, where $P_{mn}$ denotes the probability
of finding $m$ and $n$ charge $e/2$ units in the reflected and transmitted outputs, respectively.
 The cumulants are given in terms of the probabilities by
$C_1= \frac{1}{2} P_{11} + P_{02}$ and 
$C_2= \frac{1}{4} P_{11} + P_{02} - C_1^2$.
 Together with the probability normalization, it leads to
$P_{11}=4[C_1-(C_2+C_1^2)]$. Since
our results show that $C_2=C_1(1-C_1)$, this gives $P_{11}=0$, while
$P_{02}=C_1$ and $P_{20}=1-C_1$. 
Thus, the transferred charge is either $0$ or $e$,
with no intermediate $e/2$ sector.

Motivated by this result for \(q=1\), we next analyze the \(q=2\) result in terms of the three electronic-charge sectors \(P_{40},P_{22},P_{04}\) only, corresponding to transferred charges \(0,e,2e\). Again the first two cumulants are sufficient to determine the three probabilities. 
We obtain
\begin{align}
P_{04}&=\frac{C_2+C_1^2-C_1}{2},\quad
P_{22}=2C_1-(C_2+C_1^2),\\
P_{40}&=1-P_{22}-P_{04}.
\label{eq:q2probs}
\end{align}

\begin{figure}[t]
\centering
\includegraphics[width=\columnwidth]{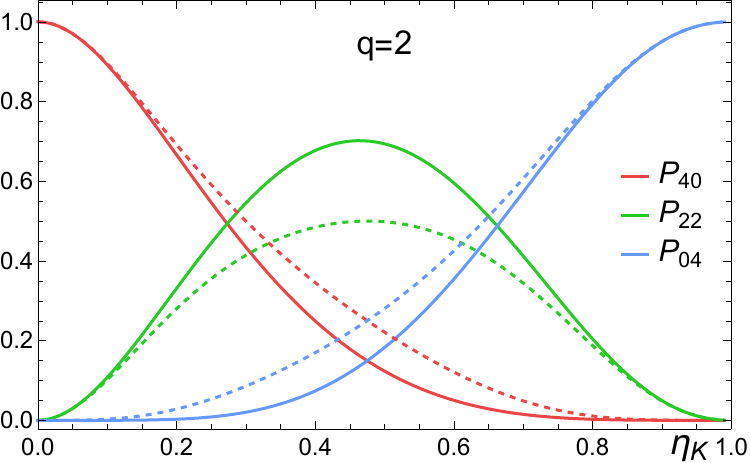}
\caption{Solid lines: probabilities $P_{04}$, $P_{22}$,
and $P_{40}$ for transferring charges $0$, $e$, and $2e$,
respectively, reconstructed from the first two cumulants
for a two-electron Leviton ($q=2$), as a function of $\eta_K= \sqrt{T_K/(2+T_K)}$. Dashed lines show the
binomial reference for two independent electrons, each transferred with probability $p=C_1/2$.}
\label{fig:probq2}
\end{figure}

Figure~\ref{fig:probq2} shows the resulting probabilities as a function
of $\eta_K= \sqrt{T_K/(2+T_K)}$, which maps the $[0,\infty]$ interval for $T_K$ to the interval $[0,1]$. The solid lines differ markedly from the
binomial reference for two independent electrons, given by
$P_{40}^{\rm bin}=(1-p)^2$, $P_{22}^{\rm bin}=2p(1-p)$, and
$P_{04}^{\rm bin}=p^2$, with $p=C_1/2$ the transfer probability of each individual electron, and shown as dashed
lines. In particular, $P_{22}$ is strongly enhanced in the crossover regime, showing a clear antibunching behavior: the interacting junction favors events in
which one electron is transferred while the other is not. Thus, the
departure from binomial partitioning is already accessible
from the average transferred charge and its fluctuations, already at the level of  two-electron states~\cite{Ubbelohde2015}.

\textit{Full counting statistics for integer Levitons.---}
The same analysis could in principle be extended to $q=3,4$ and higher values.
However, the corresponding four and five probabilities cannot
be reconstructed from $C_1$ and $C_2$ alone, while a direct
calculation of higher cumulants from expressions analogous to
Eq.~(\ref{eq:C2}) rapidly becomes cumbersome. For integer Levitons, the
full counting statistics can instead be obtained directly.

 The full counting statistics (FCS) gives the complete probability distribution
of the transferred charge. Denoting by $P_n$ the probability
of transferring $n$ electrons, its characteristic function is
$\chi(\lambda)=\sum_{n=0} P_n e^{in\lambda}$, and the
charge cumulants are obtained from derivatives of
$\ln\chi(\lambda)$.  For integer Levitons, $P_n \equiv P_{2(N-n),\,2n}$,
with $P_{m,n}$ defined in the previous section. 
For a scattering problem, the Levitov formula~\cite{Levitov1993,Klich2003} gives the characteristic function as:
\begin{equation}
\chi(\boldsymbol{\lambda}) = \mbox{det} \left[\mathbb{1}+ \hat{n} \left(S^{\dagger} e^{i \boldsymbol{\lambda}} S e^{-i \boldsymbol{\lambda}} -\mathbb{1}\right) \right],
\label{eq:chiLevitov}
\end{equation}
where $\hat n$ is the single-particle occupation operator
of the initial state, $S$ is the scattering matrix relating
incoming and outgoing states, and
$\boldsymbol{\lambda}$ is a diagonal matrix of counting
fields. In general, Eq.~(\ref{eq:chiLevitov}) involves
infinite-dimensional single-particle operators.

However, the formula simplifies when the initial state is created by an integer Lorentzian pulse, $q=N$.
Indeed the excess incoming
state is then a Slater determinant of $N$ orthonormal Leviton wave
packets $\phi_n(\omega)= \sqrt{2} \Theta(\omega) e^{-\omega} L_{n-1}(2 \omega)$ (for $\tau_0=1$),
where $L_n$ is the Laguerre polynomial~\cite{Keeling2006,Grenier2013}.
For the refermionized problem, the stationary scatterer acts
on the finite-dimensional Leviton subspace through
\begin{equation}
A_{mn}=
\int_0^\infty d\omega \,
\phi_m^*(\omega)\,\mathcal T(\omega)\,\phi_n(\omega),
\quad
\mathcal T(\omega)=\frac{T_K^2}{\omega^2+T_K^2}.
\label{eq:A}
\end{equation}
The FCS then reduces to the finite $N \times N$ determinant
\begin{equation}
\chi_N(\lambda)=
\det_N\!\left[\mathbb{1}+(e^{i\lambda}-1)A\right],
\label{eq:FCS}
\end{equation}
with $\lambda$ a simple scalar counting field.
In particular, the first two cumulants are simply
\begin{equation}
C_1=\mathrm{Tr}\,A,\qquad
C_2=\mathrm{Tr}[A(\mathbb{1}-A)].
\label{eq:FCSC2}
\end{equation}
The values of $C_1$ and $C_2$ obtained from Eq.~(\ref{eq:FCSC2}) (see Fig.~\ref{fig:C2all}) coincide with the direct evaluation
of Eqs.(\ref{eq:C1})-(\ref{eq:C2}). This agreement provides an independent
check of the two calculations. 
For the case $q=1$, $A$ reduces to a $1\times1$ matrix equal to $C_1$, which gives with Eq.~(\ref{eq:FCSC2}) $C_2= C_1(1-C_1)=C_2^{\mathrm{bin}}$
for all $T_K$.
\begin{figure*}[t]
\centering
\includegraphics[width=\textwidth]{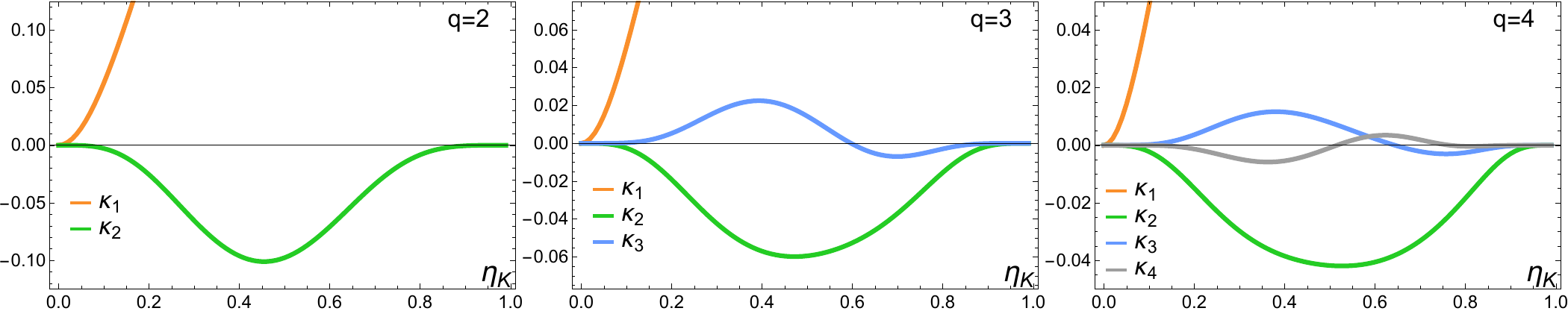}
\caption{
Symmetrized multivariate cumulants $\kappa_k$ for
the partitioning of Leviton states containing $q=2,3,4$ electrons, obtained from the full counting
statistics through Eqs.~(\ref{eq:multimoments})-(\ref{eq:multicumulants}), as a function of $\eta_K= \sqrt{T_K/(2+T_K)}$
(the vertical axis is zoomed in to resolve $\kappa_{k\ge 2}$, while $\kappa_1 = C_1/q$ rises monotonically from $0$ to $1$).
For independent and statistically equivalent electrons,
$\kappa_{k\geq2}=0$; the finite higher-order cumulants in the
crossover region therefore characterize the departure from
independent-electron partitioning.
}
\label{fig:multicumulants}
\end{figure*}

The higher cumulants for $q>2$ are easily obtained 
from the FCS. For the study of electronic correlations, it is of particular interest to define the so-called
 symmetrized multivariate cumulants $\kappa_k$, as done in Ref.~\cite{Shaju2025}. These have the important property that
for independent and statistically equivalent electrons, all the cumulants $\kappa_k$ for $k \geq 2$ identically vanish
while $\kappa_1= \frac{C_1}{N}$. The higher
multivariate cumulants therefore provide a direct measure of
the departure from independent-electron partitioning.

From the probabilities $P_n$ for transferring $n$ out of $N$ electrons,
we define the symmetrized multivariate moments
\begin{equation}
m_k=
\binom{N}{k}^{-1}
\sum_{n=k}^{N}\binom{n}{k}P_n .
\label{eq:multimoments}
\end{equation}
The corresponding cumulants $\kappa_k (k=1, \ldots,N$) are conveniently obtained from
from the generating function \(M(z)=1+\sum_{j=1}^{N}m_j z^j/j!\) at $z=0$,  with
\begin{equation}
\kappa_k=\left.\frac{d^k}{dz^k}\ln M(z)\right|_{z=0}
\qquad k=1,\ldots,N.
\label{eq:multicumulants}
\end{equation}

Figure~\ref{fig:multicumulants} shows $\kappa_k$ for
$q=2,3,4$, as a function of $\eta_K= \sqrt{T_K/(2+T_K)}$.
 Already for $q=2$, the finite negative
$\kappa_2$ directly reflects the correlated partitioning
identified from Figure~\ref{fig:probq2}. For $q=3$ and $q=4$, finite
$\kappa_3$ and $\kappa_4$ appear in the crossover regime,
showing that the FCS contains correlations beyond the
two-electron level. The results shown here are similar to the ones obtained
by Shaju et al.~\cite{Shaju2025}, revealing the similar nature of the electronic correlations in both cases:
$\kappa_2$ is negative, with a single minimum in the crossover regime, $\kappa_3$ has a positive maximum followed by a negative minimum,
while $\kappa_4$ alternates a negative, positive and negative extremum.
A notable difference with the experimental results of Shaju et al. is the observed asymmetry on Fig.~\ref{fig:multicumulants} :
 e.g., the minimum for $\kappa_3$ is less pronounced than the maximum. This
asymmetry reflects the unequal probabilities of complementary
partitions, $P_{n}\neq P_{N-n}$, arising from the energy-dependent
transmission. In the experiment of Shaju et al., it can be observed when the electron droplet is partitioned at a location where the Coulomb interaction energy differs along the two partitioning paths~\cite{SellierPrivate}.

\textit{Experimental realization.---}
\begin{figure}[t]
\centering
\includegraphics[width=7.5cm]{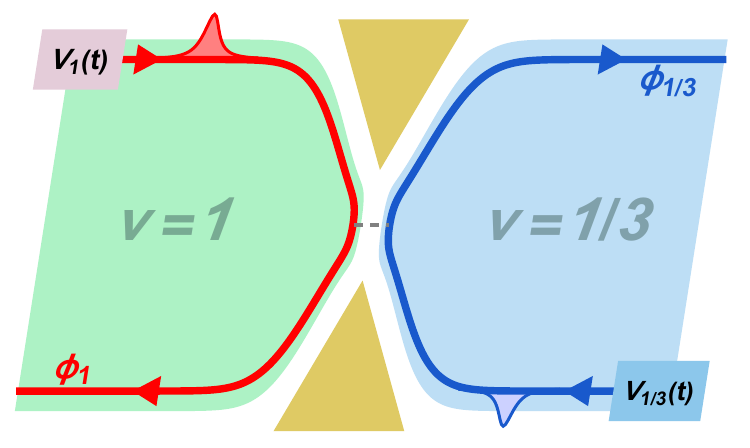}
\caption{Experimentally accessible physical realization: an inhomogeneous $(\nu_1,\nu_2)=(1,1/3)$ quantum Hall junction driven by $V_1(t)$ and $V_{1/3}(t)$ [Eq.~(\ref{eq:V1V13})] is equivalent to the $g=1/2$ effective system of Fig.~\ref{fig:setup}.}
\label{fig:setupinHom}
\end{figure}
A homogeneous FQH junction at filling factor $g=1/2$ does not correspond to a Laughlin state, but an effective $g=1/2$ junction can be realized in an inhomogeneous $(\nu_1,\nu_2)=(1,1/3)$ quantum Hall junction~\cite{Sandler1999,Hashisaka2021,Cohen2023}, as sketched in Fig.~\ref{fig:setupinHom}. In terms of canonically normalized bosonic fields $\phi_1$ and $\phi_{1/3}$, the two physical edges carry charge densities $\rho_\nu(x) = \frac{e\sqrt{\nu}}{2\pi}\partial_x\phi_\nu(x)$ ($\nu=1,1/3$) and are governed by the free Hamiltonian $H_0 = \frac{v_F}{4\pi}\sum_{\nu=1,1/3}\int dx\,[\partial_x\phi_\nu(x)]^2$. Following Sandler \textit{et al.}~\cite{Sandler1999}, the orthogonal rotation
\begin{align}
\begin{pmatrix} \phi_+ \\ \phi_- \end{pmatrix}
&= \frac{1}{\sqrt{\nu_1+\nu_2}}
\begin{pmatrix} \sqrt{\nu_1} & \sqrt{\nu_2} \\ \sqrt{\nu_2} & -\sqrt{\nu_1} \end{pmatrix}
\begin{pmatrix} \phi_1 \\ \phi_{1/3} \end{pmatrix} \nonumber \\
&=
\begin{pmatrix} \frac{\sqrt{3}}{2}\phi_1 + \frac{1}{2}\phi_{1/3} \\[4pt] \frac{1}{2}\phi_1 - \frac{\sqrt{3}}{2}\phi_{1/3} \end{pmatrix}
\label{eq:rot_Sandler}
\end{align}
separates the conserved total-charge mode $\phi_+$ ($\partial_x\phi_+ \propto \rho_1 + \rho_{1/3}$), which decouples from the QPC, from the relative mode $\phi_-$ that governs tunneling at $x=0$. Combined with the duality transformation of Ref.~\cite{Sandler1999} relating weak electron tunneling ($\propto e^{i 2\phi_-(0)}$) to weak quasiparticle backscattering, the inhomogeneous junction maps directly onto Eqs.~(\ref{eq:H0})--(\ref{eq:HT}) for a homogeneous junction with effective parameter
\begin{equation}
g = \frac{2\nu_1\nu_2}{\nu_1+\nu_2} = \frac{1}{2}.
\end{equation}
Through $\phi_{L,R} = (\phi_+ \pm \phi_-)/\sqrt{2}$, a one-sided drive $V_L(t)$ on the left effective edge (with $V_R(t)=0$) is then realized by applying
\begin{equation}
V_1(t) = \frac{\sqrt{3}+1}{4}V_L(t),
\qquad
V_{1/3}(t) = (-\sqrt{3})\frac{\sqrt{3}-1}{4}V_L(t)
\label{eq:V1V13}
\end{equation}
to the physical $\nu=1$ and $\nu=1/3$ edges, respectively. The transferred charge is invariant under this mapping, so that the cumulants calculated above are directly accessible in the physical junction.

\textit{Conclusion. ---}
In summary, we have obtained a nonperturbative description of the
transferred charge and its fluctuations at the exactly solvable
effective Luttinger parameter $g=1/2$, for an arbitrary voltage
drive. For integer-charge Lorentzian pulses, the full counting
statistics reduces to a finite-dimensional determinant and contains
only integer-electron transfer sectors. For $N=2$, the first two
cumulants determine the partition probabilities and reveal a clear
departure from the binomial reference for independent, equivalent
electrons. For $N=3$ and $4$, the symmetrized cumulants constructed
from the FCS quantify higher-order departures from this reference.
The mapping to an inhomogeneous $(\nu_1,\nu_2)=(1,1/3)$ junction
offers a route to testing these predictions experimentally. 
An extension to homogeneous quantum point contacts
in Laughlin states would be particularly appealing~\cite{Bertin2024}.

\begin{acknowledgements}
{We thank H. Sellier for enlightening exchanges on the partitioning experiment~\cite{Shaju2025}, and G. F\`eve, G. M\'enard, F. Parmentier 
and C. B\"auerle for useful discussions.
We  acknowledge  funding from  the  Agence  Nationale  de  la  Recherche  under  the France 2030 programme, reference ANR-22-PETQ-0012,
and from the ERC advanced grant “ASTEC” (Grant No.101096610).}
\end{acknowledgements}

\section*{END MATTER}

\textit{Refermionization and arbitrary-drive correlators.}--- At $g = 1/2$, the boundary vertex operator in Eq.~(\ref{eq:HT}) has scaling dimension $1/2$ and can be refermionized in terms of a chiral Dirac fermion $\psi(x) = \frac{1}{\sqrt{2\pi a}} e^{i\phi_-(x)}$ together with a local Majorana mode $\gamma = \gamma^\dagger$ ($\gamma^2 = 1/2$) ensuring anticommutation~\cite{Chamon1995,Sandler1999}. Setting $\hbar = v_F = 1$ (with the short-distance cutoff absorbed in $\Gamma$) and choosing a gauge where $\Gamma$ is real, the antisymmetric sector is governed by the quadratic Hamiltonian
\begin{equation}
H_{\mathrm{ref}} = -i \int dx\, \psi^\dagger(x)\partial_x\psi(x) + \sqrt{2}\Gamma\, \gamma \big[\psi(0) - \psi^\dagger(0)\big].
\label{eq:Href}
\end{equation}
Integrating the linear Heisenberg equations of motion $\partial_t \psi(x,t) = -\partial_x \psi(x,t) - i\sqrt{2}\Gamma\gamma(t)\delta(x)$ and $\dot{\gamma}(t) = -i\sqrt{2}\Gamma[\psi(0,t) - \psi^\dagger(0,t)]$ across $x=0$ relates $\psi_{\mathrm{out}}(t) \equiv \psi(0^+,t)$ to $\psi_{\mathrm{in}}(t) \equiv \psi(0^-,t)$ as
\begin{equation}
\psi_{\mathrm{out}}(t) = \psi_{\mathrm{in}}(t) - T_K \int_{-\infty}^t dt'\, e^{-T_K(t-t')} \big[\psi_{\mathrm{in}}(t') - \psi_{\mathrm{in}}^\dagger(t')\big],
\label{eq:psi_out_t}
\end{equation}
where $T_K = 2|\Gamma|^2$ is the crossover energy scale, which directly yields Eq.~(\ref{eq:scattering}) in frequency space.

At zero temperature, the two independent equilibrium contractions of the incoming field are
\begin{align}
\label{eq:freeG}
G_0^<(\tau) &= \langle \psi_{\mathrm{in}}^{(0)\dagger}(t')\psi_{\mathrm{in}}^{(0)}(t)\rangle = \frac{1}{2\pi}\frac{1}{t_0 + i\tau}, \\
G_0^>(\tau) &= \big[G_0^<(\tau)\big]^*,
\end{align}
with $\tau = t - t'$ and $t_0$ the short-time cutoff. Under an arbitrary voltage drive $V(t)$, each incoming field acquires the Peierls phase $\varphi(t)$ of Eqs.~(\ref{eq:drive})--(\ref{eq:drivencorr}), while anomalous incoming contractions $\langle \psi_{\mathrm{in}}(t')\psi_{\mathrm{in}}(t)\rangle_V$ vanish. Expressing the current operator $I(t) = \frac{e}{2}[\psi_{\mathrm{in}}^\dagger(t)\psi_{\mathrm{in}}(t) - \psi_{\mathrm{out}}^\dagger(t)\psi_{\mathrm{out}}(t)]$ in terms of $\psi_{\mathrm{in}}, \psi_{\mathrm{in}}^\dagger$ via Eq.~(\ref{eq:psi_out_t}) and taking the expectation value gives Eq.~(\ref{eq:Iarbitrary}). Applying Wick's theorem to the connected correlator $S(t_1,t_2) \equiv \langle \delta I(t_1)\delta I(t_2)\rangle$ yields the exact zero-temperature noise kernel
\begin{widetext}
\begin{align}
S(t_1,t_2) = \frac{e^2 T_K^2}{(2\pi)^2} \int_0^\infty d\tau_1 \int_0^\infty d\tau_2\, e^{-T_K(\tau_1+\tau_2)} \Bigg[ &-\sin\big(\varphi(t_1)-\varphi(t_2-\tau_2)\big)\sin\big(\varphi(t_2)-\varphi(t_1-\tau_1)\big) \nonumber\\
&\times \frac{(t_1-t_2+\tau_2)+it_0}{(t_1-t_2+\tau_2)^2+t_0^2} \frac{(t_2-t_1+\tau_1)-it_0}{(t_2-t_1+\tau_1)^2+t_0^2} \nonumber\\
+\, \Big[\cos\big(\varphi(t_1)-\varphi(t_2)\big)\cos\big(\varphi(t_2-\tau_2)&-\varphi(t_1-\tau_1)\big) - 1\Big] \frac{(t_1-t_2)+it_0}{(t_1-t_2)^2+t_0^2} \frac{(t_2-\tau_2-t_1+\tau_1)-it_0}{(t_2-\tau_2-t_1+\tau_1)^2+t_0^2} \Bigg].
\label{eq:Sexplicit}
\end{align}
\end{widetext}
Together with Eq.~(\ref{eq:C2}), this expression determines the charge fluctuations $C_2$ for an arbitrary pulse $V(t)$ solely in terms of the single scale $T_K$.

\textit{Finite-rank reduction for integer Levitons.}--- Because Eq.~(\ref{eq:scattering}) couples $\psi(\omega)$ to $\psi^\dagger(-\omega)$, it is convenient to restrict frequencies to $\omega > 0$ and introduce the two-component Nambu vector $\Psi_\alpha(\omega) = \big(c_\alpha(\omega),\, d_\alpha(\omega)\big)^T \equiv \big(\psi_\alpha(\omega),\, \psi_\alpha^\dagger(-\omega)\big)^T$ for $\alpha \in \{\mathrm{in}, \mathrm{out}\}$, where $c_\alpha(\omega)$ and $d_\alpha(\omega)$ obey standard canonical anticommutation relations on $\omega > 0$. The scattering relation~(\ref{eq:scattering}) then takes the unitary two-channel form
\begin{equation}
\Psi_{\mathrm{out}}(\omega) = S(\omega)\Psi_{\mathrm{in}}(\omega), \quad S(\omega) = \frac{1}{\omega + iT_K} \begin{pmatrix} \omega & iT_K \\ iT_K & \omega \end{pmatrix}.
\label{eq:NambuS}
\end{equation}
In terms of the normal-ordered densities at $\omega > 0$, $\int_{-\infty}^\infty \frac{d\omega}{2\pi} {:}\psi_\alpha^\dagger(\omega)\psi_\alpha(\omega){:} = \int_0^\infty \frac{d\omega}{2\pi} \big[c_\alpha^\dagger(\omega)c_\alpha(\omega) - d_\alpha^\dagger(\omega)d_\alpha(\omega)\big]$. Since $S(\omega)$ conserves $c^\dagger(\omega)c(\omega) + d^\dagger(\omega)d(\omega)$, the transferred charge operator~(\ref{eq:charge}) reduces to
\begin{equation}
Q = \int_0^\infty \frac{d\omega}{2\pi}\, \big[d_{\mathrm{out}}^\dagger(\omega)d_{\mathrm{out}}(\omega) - d_{\mathrm{in}}^\dagger(\omega)d_{\mathrm{in}}(\omega)\big],
\label{eq:Q_d}
\end{equation}
so that the transferred charge simply counts fermions scattered into the $d_{\mathrm{out}}$ channel. For a Lorentzian pulse carrying an integer charge $q = N$, the excess initial state contains no hole excitations ($d_{\mathrm{in}}(\omega)|N\rangle = 0$ for $\omega > 0$) and is a pure Slater determinant of $N$ orthonormal positive-energy Leviton wave packets $\{\phi_n(\omega)\}_{n=1}^N$ in the $c_{\mathrm{in}}$ channel~\cite{Keeling2006,Grenier2013}. In the $(c,d)$ basis, the occupation operator $\hat{n}$ and the counting-field matrix $e^{i\boldsymbol{\lambda}}$ therefore read
\begin{equation}
\hat{n} = \begin{pmatrix} \hat{P}_N & 0 \\ 0 & 0 \end{pmatrix},
\qquad
e^{i\boldsymbol{\lambda}} = \begin{pmatrix} 1 & 0 \\ 0 & e^{i\lambda} \end{pmatrix},
\end{equation}
where $\hat{P}_N = \sum_{n=1}^N |\phi_n\rangle\langle\phi_n|$ is the rank-$N$ projector onto the Leviton subspace. Only the $(1,1)$ block of $S^\dagger(\omega) e^{i\boldsymbol{\lambda}} S(\omega) e^{-i\boldsymbol{\lambda}} - \mathbb{1}$, equal to $(e^{i\lambda}-1)|S_{21}(\omega)|^2 = (e^{i\lambda}-1)\mathcal{T}(\omega)$, then contributes to the Levitov-Lesovik determinant~(\ref{eq:chiLevitov}), which directly reduces on the $N$-dimensional subspace of $\hat{P}_N$ to Eqs.~(\ref{eq:A})--(\ref{eq:FCS}).

%%% equations for the talk %%%

\end{document}